# Review of Results from the FN-II Dense Plasma Focus Machine

**J.J.E. Herrera, F. Castillo, I. Gamboa, J. Rangel**
*Instituto de Ciencias Nucleares, Universidad Nacional Autónoma de México*
*A.P. 70-543, Ciudad Universitaria, 04511, México, D.F., México*

**J.I. Golzarri and G. Espinosa**
*Instituto de Física, Universidad Nacional Autónoma de México.*
*Apdo. Postal 20-364, 01000 México, D.F. México*

**Abstract**
The FN-II is a small dense plasma focus (4.8 kJ at 36 kV), operated at the Universidad Nacional Autónoma de México. Substantial effort has been dedicated to the study of the anisotropy in the neutron and hard X-ray radiation. Concerning the former, it has been observed that there is an anisotropic distribution superposed on a far larger isotropic one. These clearly separated effects can be interpreted as the consequence of two different neutron emission mechanisms. The angular distribution of hard X-rays and ions is also studied within the chamber with TLD and CR-39 detectors respectively. Two maxima are found around the axis of the device for X rays within the 20-200 keV range.

# 1. Introduction

The dense plasma focus device has been recognized as an interesting radiation source, and specially as a plasma based neutron source (via $^2H(^2H,n)^3He$ reactions), since its conception in the early days of fusion research (Mather 1971). In spite of the difficulties it presents, mostly regarding the poor uniformity of the radiation yield from shot to shot, it has prevailed, particularly in developing countries, due to the simplicity of its engineering, its low cost, as compared to other fusion research devices, and the wealth of its phenomena (Brzosco et al. 1997, Bernard et al. 1998). Up to date, it is arguably the most accessible source of fusion neutrons, and research is underway in order to produce small, high repetition rate devices for applications, not only as neutron sources (Moreno et al. 2002), but also as X-ray sources for microelectronics lithography (Lee et al., 1998) and biomedical studies (Dubrovsky et al., 1998). However, evidence exists since the early days of plasma focus research that the origin of such neutrons is not necessarily thermonuclear (Bernstein and Hai 1970, Milanese and Pouzo 1978). Ever since, anisotropy of the neutron fluence was observed (Lee et al. 1972), and the actual mechanism of neutron production in plasma foci has remained a source of controversy.

It has been well established that an energetic ion beam is produced in the forward direction (Jäger and Herold, 1987, Rhee et al. 1988, Moo et al., 1991, Kelly and Márquez, 1996), which originates the anisotropic neutron generation (Aliaga and Choi 1998, Castillo et al. 1997, 2002). Although there has been considerable theoretical work on the study of the acceleration mechanisms for such ion beams, proper diagnostics that can discriminate between existing theories, are still lacking. The dynamics of the ions is rather complex during the compression phase, and the rise time of neutron signals is faster than the thermalisation time. Thus, even if one surmises that thermalisation might be achieved within the time scale of this phase, it is reasonable to believe that different neutron generation mechanisms coexist, yielding both isotropic and anisotrpic components in the neutron emission (Castillo et al. 2000, 2002, 2003). On the other hand, many laboratories rely on the measurements of calibrated activation counters in order to estimate the total neutron yield, but if the neutron emission is anisotropic, their results can be affected by the position of such counters. Placing counters both head on and side on may help to get a better picture of the neutron emission, but as shown by Castillo et al. (2003), this is not enough. It was shown that there are both isotropic and anisotropic components of the neutron angular distribution, and that in the two machines studied (the FN-II in Mexico City, and the PACO in Tandil, Argentina), the anisotropic component is smaller than the isotropic one. The case of PACO is particularly interesting and instructive, because although its anisotropy is larger than usual, it was found that the neutron fluence close to the axis is so narrow, that it only accounts for a small fraction of the total neutron yield ($\leq 5\%$). Thus, it follows that the fact that the ratio of the neutron fluence on axis to the fluence at $90^o$ is large, does not necessarily mean that the beam-target mechanism is responsible for most of the neutrons.

Time resolved measurements of the evolution of the neutron emission for angles between $0^o$ and $90^o$ from the axis, have been made by Aliaga-Rossel and Choi (1998) for a medium size device (DPF-78, 28 kJ, 60kV) and by Yap et al. (1997) in a small device (3.3 kJ, 15 kV). They have shown that the head-on neutron pulse lasts longer than the side-on pulse. Furthermore, they have found that, when a doping gas is added, the anisotropy is significantly increased; the neutron pulse decays faster at larger angles, while the head-on pulse remains essentially unchanged. As the concentration of the doping gas increases, the length of the side-on pulse decreases, and a double structure in the neutron emission is revealed. If the Z of the doping gas is increased, the second peak disappears from the side-on signals. This suggests the existence of two different neutron generation mechanisms, which can be selectively affected by the doping level, as well as the Z of the doping gas. This might be related to earlier observations in larger devices, such as those at the Frascati (1MJ) (Maisonnier et al., 1972) and Poseidon (780 kJ, 60 kV) (Herold et al., 1985, Jäger and Herold 1987), devices. In both of them, neutrons were emitted in two distinct periods; the first one during the final phase of radial compression, and the second one during the expansion of the plasma column. However, it is not clear, so far, if the mechanisms responsible for the two neutron pulses in small devices are the same as those involved in large ones, since in the latter the size of the second pulse is clearly distinguished and larger than the first one. In contrast, the second pulse in medium and small size devices is smaller than the first one, and practically masked by it.

Although they fail to give information on the plasma dynamics, time integrated diagnostics are also useful, particularly when statistical information, such as the average X-ray or neutron yield, are needed. This would be the case when applications are in mind, where a sample is subject to radiation from several shots. In Castillo et al. (2003) we measured the angular distribution of the neutron yield, outside the vacuum chamber, 1 m away from the plasma column, using plastic track detectors, between $-170^o$ and $170^o$. In this work we use the same kind of detectors in order to obtain angular distributions for neutrons and protons (resulting from $^2H(^2H,p)^3H$ reactions), within the chamber, at a 13 cm. radius, using a semicircular holder. The hard X-ray angular distribution is measured using TLD-200 dosimeters. However, in this work we are physically constrained to measure the angular distribution within the $\pm 70^o$ range.

In section 2 we describe the experimental ser-up. The methods used to obtain the angular distributions for neutrons and protons are discussed in section 3, and the measurement of the angular distribution of X-rays is shown in section 4. Section 5 is devoted to conclusions.

## 2. Experimental set-up

This work was performed using the FN-II small plasma focus device, operated at the Instituto de Ciencias Nucleares, UNAM (Castillo et al., 2002). It is a 5 kJ device at 37 kV, with an oxygen-free copper anode, 40 mm long, with a 50 mm diameter. The co-axial cathode is formed by ten copper rods arranged in a squirrel cage configuration at a radius of 50 mm. The insulator is an annular *Pyrex®* tube located at the base of the anode, matching its diameter. The energy storage is provided by four 1.863 µF capacitors in parallel, and the discharge is triggered by a simple mushroom electrode spark gap. Throughout the present work the plasma focus was operated in its neutron optimised regime, corresponding to a 2.75 torr deuterium gas pressure. In this regime a peak focus current of 350 kA and an average neutron yield of $\sim 3 \times 10^8$ per shot are obtained. The evolution of the current derivative is obtained with a Rogowski coil. Two silver foil activation counters, one on axis, and the other at $90^o$, were used to measure the neutron yield.

Neutron and proton CR-39 detectors, along with thermoluminescent TLD-200 dosimeters, were placed on a 13 cm diameter semicircular holder inside the chamber around the central electrode of the plasma focus (Fig. 1.), as will be explained in next two sections.

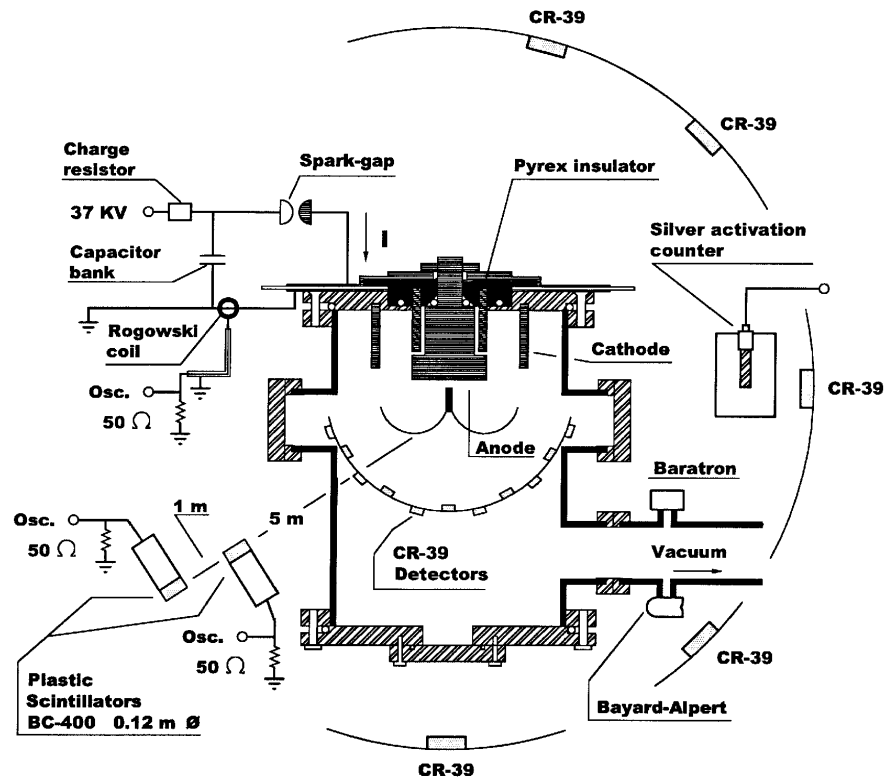

Figure 1. Scheme of the *Fuego Nuevo II* PF-device. Also shown is the diagnostics equipment used for ion and neutron measurements.

## 3. Angular distribution of protons and neutrons

Plastic nuclear track detectors have played an important role in plasma focus research. They have been used in diagnostics of the axially accelerated ion beam (Sadowski et al. 1998, 2000, Skladnik-Sadowska et al. 1999, 2001), as films of Thompson spectrometers (Rhee 1985), including a time resolved version (Rhee 1987 and 1988). They have also been used for the detection of charged reaction products (Al-Mashhadni et al. 1994, Todorovick 1995), and in some cases, images of the charged particle emission sources have been obtained using the *camara obscura* technique (Jäger et. al 1985, Jäger and Herold 1987, Zakahulla 1999, Springham 2002). The way nuclear track detectors work can be summarised as follows. When a charged particle is slowed down in the detector, mainly by inelastic interactions with the plastic material, it leaves latent tracks of broken molecular bonds. These tracks can then be exposed, by attacking the plastic with some chemical agent, such as NaOH or KOH. Since the etch rate is larger over the track of broken bonds, a conical pit is produced, whose diameter depends, for a given etching time, on the charge and energy of the particle that produced it. Furthermore, the ellipticity of the pit give information about the angle of incidence of the charged particle. It can be easily observed with an optical microscope. Electrochemical etching can also be used, where the etched tracks filled with the conducting etchant act as needle electrodes (Brede 1999). In this case, the electric field is amplified at the tip of the cone, and if it is high enough, electric breakdown occurs, producing an enlargement of the track. However, we have only used chemical etching in our work.

The use of CR-39 ($C_{12}H_{16}O_7$) nuclear track detectors for proton detection is straightforward. Their use in neutron dosimetry was reviewed by Matihulla et al. (1990). Their use as a diagnostic for neutrons from deuterium-deuterium (D-D) and deuterium-tritium (D-T) reactions has been studied by Collopy et al (1992), using an accelerator, and more recently by Frenje et al. (2002), using the OMEGA laser facility. In the case of D-D reactions, with which we are concerned, 2.45 MeV neutrons are produced, which interact with the detector mainly through elastic (n,p) scattering of protons form the detector material. Radiator foils, such as polyethylene, can be placed in front of them, in order to increase the detection efficiency. At higher energies of the neutrons, even carbon and oxygen scattered nuclei can produce etchable tracks.

The proton detectors, with dimensions 1.8 x 0.9 cm$^2$, 500 μm thick, were placed on the top side of the holder, covered with 100 μm Al filters, which are able let through the 2.8 MeV protons from the fusion reactions, and stop other charged reaction products, such as 1 MeV tritons and 0.8 MeV $^3$He. They should also be able to stop the lower energy deuterons that are accelerated in the discharge, away from the axis, and impurity ions which result from the erosion of the electrode. Indeed, they can still detect neutrons which result from n-p reactions within the detector. For this reason, the analysis is made over the top surface of the chips. The neutron detectors are placed on the bottom side of the *Teflon* holder, which is 1 cm. thick. Additional 1mm. thick polycarbonate sheets were placed in front of the neutron detectors, for an enhanced n-p reaction rate. When in our earlier work, the neutron angular distribution was measured 1m away from the focus, outside the chamber (Castillo et al, 2003), we had the advantage of being able to treat the plasma column more as a point-like source, and measuring over a wider angular range, but at the cost of neutron fluence, and consequently a larger number of shots were necessary. On the other hand, charged particle diagnostics can only be possible inside the chamber. The detectors were placed at 0º, ±15º, ±40º and ±70º, and were exposed to 50 shots.

After exposure the nuclear track detectors were etched by the standard procedure, in KOH solution at controlled temperature ( 60 ± 1 °C ). The track density and track diameter distribution were measured with a digital image analysis system (Gammage and Espinosa, 1997). The proton detector placed at 0º deserves special attention, since it shows a 5mm. diameter blotch, where an extremely high density of tracks is observed. This is due to the high energy deuteron beam accelerated by the plasma column. However, outside this blotch, it is still possible to measure the track density. The number of tracks per cm$^2$, counting only the circular ones, at each angle, is shown in Figure 2. These results were obtained with 12 hours of etching. Their average diameter is 3 ± 1 μm., which shows that suggests a mono-energetic distribution of a single kind of particles. The angular distribution for the detectors below the holder, showing roughly a tenth of the density as those above the holder, is shown in Figure 3. In this case 16 hours of etching were necessary. In both cases the maximum is observed close to the axis, but slightly shifted in the positive direction.

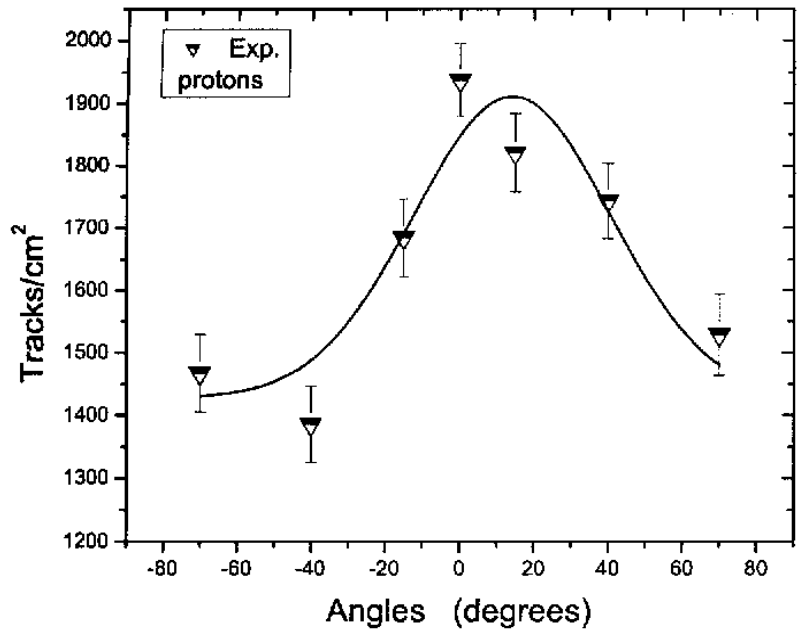

Figure 2. Angular distribution of the track density for the detectors on top of the holder. Both proton and neutron tracks are included.

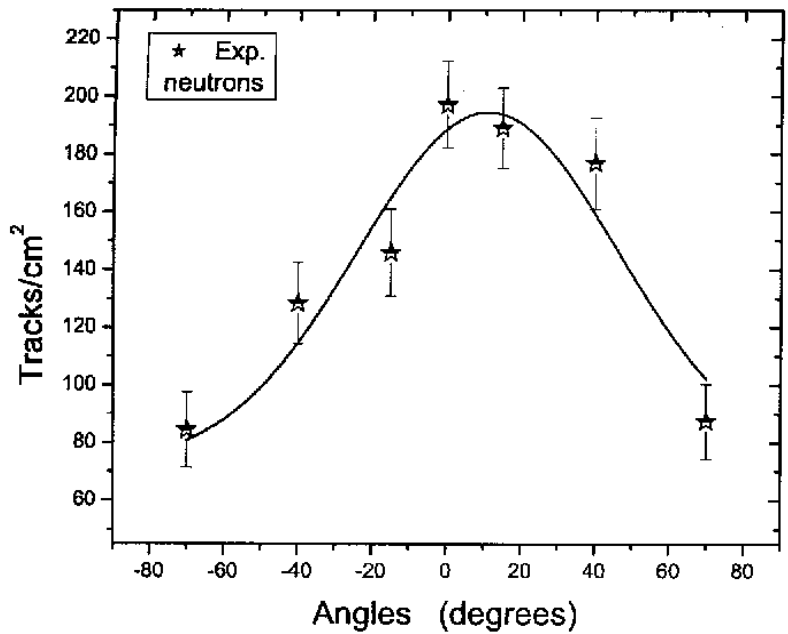

Figure 3. Angular distribution of the track density for the detectors below the holder. Only neutrons can be detected in this case.

A reasonable doubt can be cast regarding the fact that the detectors on top of the holder can detect both neutrons and protons. For this reason, it is necessary to obtain the neutron distribution simultaneously, in order to compensate this effect. A comparison between figures 2 and 3 suggests that the tracks due to n-p reactions could at most be a tenth of those observed on top, since all other charged particles are stopped by the holder.

**4. X-ray angular distribution**

The angular distribution of X- rays can be obtained by means of thermoluminescent TLD-200 dosimeters (McKeever et al. 1995). In this case, radiation produces excited states within the solid, which can be released by heating. The luminescent response is proportional to the radiation received by the dosimeter, Their sensitivity depends on the material used. The TD-200 is made of $CaF_2:Dy$, and is sensitive to rays within the 20-200 keV range, with a pronounced peak at 30 keV.

Like the proton CR-39 detectors, they were placed on top of the holder, covered by aluminium foil, in order to protect them from the plasma discharge, at $0°$, $±10°$, $±20°$, $±30°$, $±40°$, $±50°$, $±60°$, and $±70°$. Figure 4 shows the resulting angular distribution, where statistics of 10 batches of dosimeters have been accumulated.

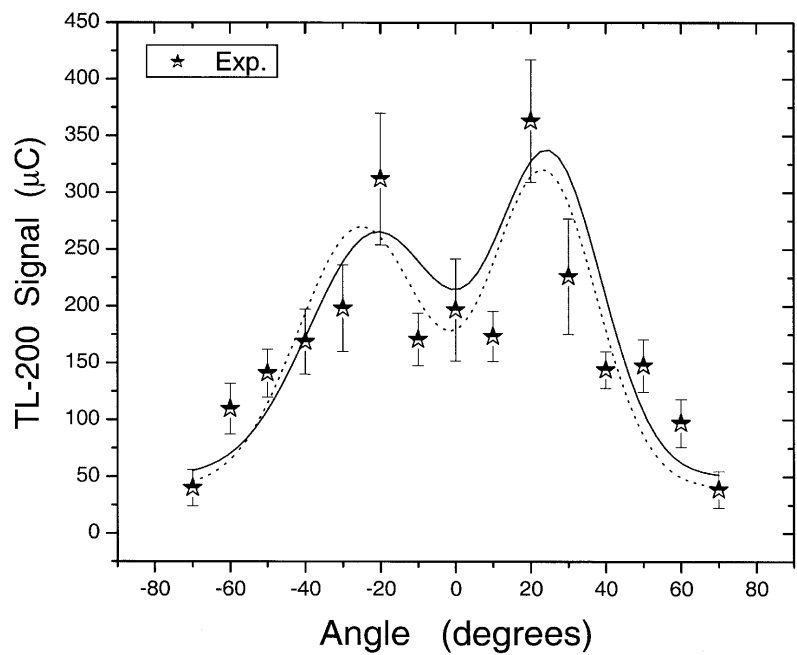

Figure 4. Angular distribution of X-rays in the 20-200 keV range by means of TLD-200 ($CaF_2:Dy$) thermoluminescent dosimeters. The two curves are possible fits to the experimental data.

It is interesting to note that, opposed to the proton and neutron angular distributions, the X-ray distribution is bi-modal, peaked at $±20°$. This apparently happens because the source of this radiation is the electron beam that is accelerated towards the electrode. This is interesting for applications purposes, since it tells us that the axis is not the most illuminated position.

## 5. Conclusions

Angular distributions of protons, neutrons and X-rays, within the range of 20-200 keV have been obtained for the small FN-II dense plasma focus device, within the discharge chamber. The neutron distribution shows the essentially the same features obtained in our previous work (Castillo 2003), in which we showed that the neutron fluence is composed of an anisotropic contribution mounted on an isotropic one. The shape of the proton distribution is very similar to the neutron one. This simultaneous measurement is necessary, in order to estimate the compensation necessary in order to obtain the proton distribution. It is found in this work that the neutrons can account for at most for one tenth of the observed density. The nature of the array unfortunately forbids the possibility of exploring the yield at angles beyond $90^o$.

Although the well collimated deuteron beam, accelerated by the plasma column, was clearly observed on the $0^o$ detector on top of the holder, its study was beyond the scope of this work, so it was not further studied, but there is indeed a potentiality for the use of this kind of detectors for such purpose. This has indeed been explored by other authors in the past (Springham,2002).

It was also found, by means of TLD-200 thermoluminescent dosimeters, that the X-rays, within the 20-200 keV range show a bimodal angular distribution peaked at $\pm 20^o$; a natural consequence of he fact that they are produced by the collimated electron beam which is accelerated along the axis towards the electrode.

**Acknowledgments**

This work was partially supported by the DGAPA-UNAM grant IN105100 and CONACYT grant 27974-E.